\documentclass[1p]{elsarticle}

\usepackage{hyperref,units}

\usepackage{units} 

\providecommand{\bc}{\begin{center}}
\providecommand{\ec}{\end{center}}
\providecommand{\be}{\begin{equation}}
\providecommand{\ee}{\end{equation}}
\providecommand{\bea}{\begin{eqnarray}}
\providecommand{\eea}{\end{eqnarray}}
\providecommand{\bdm}{\begin{displaymath}}
\providecommand{\edm}{\end{displaymath}}
\providecommand{\bdma}{\begin{eqnarray*}}
\providecommand{\edma}{\end{eqnarray*}}
\providecommand{\ba}{\begin{eqnarray*}}
\providecommand{\ea}{\end{eqnarray*}}
\providecommand{\bi}{\begin{itemize}}
\providecommand{\ei}{\end{itemize}}
\providecommand{\benum}{\begin{enumerate}}
\providecommand{\eenum}{\end{enumerate}}

\providecommand{\refkl}[1]{(\ref{#1})}

\providecommand{\twoCases}[4]{
  \left\{ 
    \begin{array}{ll} 
      #1 & #2 \\
      #3 & #4 
    \end{array} 
  \right.
}

\providecommand{\text}[1]{{\mbox{ #1}}}

\usepackage{tikz}

\providecommand{\fig}[2]{
   \begin{center}
     \includegraphics[width=#1]{#2}
   \end{center}
}

\providecommand{\diff}[1]{ \ {\rm d} #1} 
\providecommand{\ablpart}[2]{\frac{\partial #1}{\partial #2}}  
\providecommand{\abl}[2]{\frac{{\rm d} #1}{{\rm d} #2}}  

\providecommand{\sub}[1]{_{\rm #1}}
\renewcommand{\sup}[1]{^{\rm #1}}

\providecommand{\vecbeta}{\vec{\beta}}

\journal{Physica A}

\begin{document}

\begin{frontmatter}

\title{Self-Driven Particle Model for Mixed Traffic and Other
Disordered Flows}

\author[mymainaddress]{Venkatesan Kanagaraj}
       \ead{vkanagaraj.iitm@gmail.com}

\author[mymainaddress]{Martin Treiber\corref{mycorrespondingauthor}}
\cortext[mycorrespondingauthor]{Corresponding author}
\ead{martin@mtreiber.de}

\address[mymainaddress]{Technical University of Dresden, Germany}

\begin{abstract}
Vehicles in developing countries have widely varying dimensions and speeds, and drivers
tend to not follow lane discipline. In this flow state called ``mixed
traffic'', the interactions 
between drivers and the resulting maneuvers resemble more that of general disordered
self-driven particle systems than that of the orderly lane-based
traffic flow of industrialized countries. We propose a general
multi particle model for such self-driven ``high-speed particles'' and 
show that it reproduces the observed characteristics of mixed
traffic. The main idea is 
to generalize a conventional acceleration-based car-following model to
a two-dimensional 
force field. For in-line following, the model reverts to the
underlying car-following model, for very slow speeds, it reverts to an
anisotropic social-force model for pedestrians. 
With additional floor fields at the position of lane markings, the
model reverts to an 
integrated car-following and lane-changing model with continuous lateral dynamics
including cooperative aspects such as zip merging. With an adaptive cruise control (ACC)
system as underlying car-following model, it becomes a controller for
the acceleration and steering of autonomous vehicles in mixed or lane-based traffic. 
\end{abstract}


\end{frontmatter}



\section{Introduction}
\begin{figure}
\fig{0.9\textwidth}{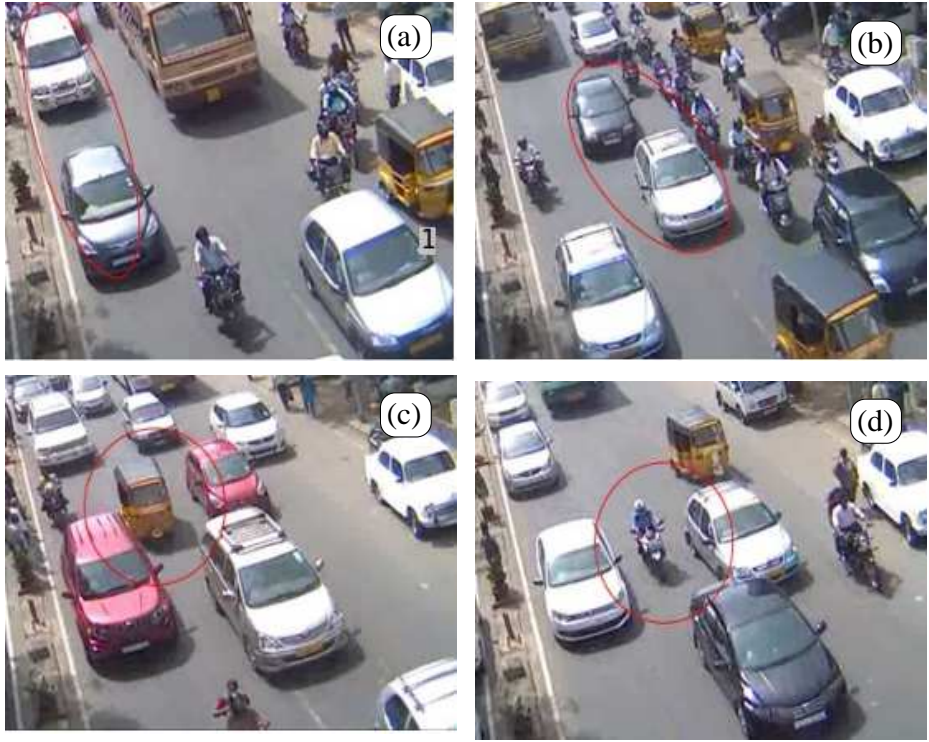}
\caption{\label{fig:behavior}The four driving behavioral patterns of
  mixed traffic. (a) In-line car following, (b) staggered 
following, (c) following between two vehicles, and (d) passing.}
\end{figure}

Traffic flow in developing countries is growing disproportionately
fast now accounting for 80\% of the total road accidents and an
estimated economic loss of 1-2\% of the GNP [1]. 
As visualized in Fig.~\ref{fig:behavior}, vehicles in developing
countries have widely varying dimensions and speeds, and drivers tend
to not follow lane discipline. In addition to in-line car following
(Fig. 1a), a model for mixed traffic should describe staggered
following (Fig. 1b), follow 
ing between two vehicles (Fig. 1c) and passing
(Fig. 1d)~\cite{Robertson2002motorcycling,VenkatTRC2015}. 
Generally, the
interactions between drivers and the resulting maneuvers in this flow
state, called ``mixed 
traffic'', can only be described fully two-dimensionally. In a wider
context, mixed traffic becomes  also increasingly relevant in
industrialized countries: On the one hand, bicycle traffic and its
interaction with driving and standing cars and pedestrians has the
attributes of mixed traffic, particularly on bike lanes allowing
several cyclists to drive in parallel while only single-file bicycle
traffic has been modelled~\cite{Andresen2014_universal,hoogendoorn2016bicycle}. On the
other hand, the concept of "shared space" creates mixed traffic (of
motorized vehicles, bicycles, and pedestrians) by
design~\cite{Schoenauer-sharedSpace}. 

The core behavioral models such as car-following models
\cite{Gipps81,Krauss-traff98,Bando-jphys,Opus} and lane-changing
models
\cite{Gipps86,Hidas-02,MOBIL-TRR07,Moridpour-reviewLC,TreiberKesting-Book,Chowdhury-reviewLC}
are designed for
lane-based traffic,only. 
There exist a few models describing staggered car following
\cite{Lee2009new,Jin2012,Gunay2007}, or continuous lane changing based on a 
constant lateral speed~\cite{Oketch2000}. To our knowledge,
there are no models describing the full dynamics of non-lane based
traffic. Existing self-driven particle models like the social-force
model for pedestrians do not apply because, 
due to the kinematic constraints of high speeds and the negative
consequences of crashes, 
drivers behave qualitatively differently as pedestrians or point-like
particles. 

In this contribution, we propose a continuous, fully two-dimensional
microscopic model for motorized and non-motorized vehicles in
unidirectional traffic and show that it reproduces the observed
characteristics of mixed traffic on arterials in an Indian city. The
main idea is to generalize conventional acceleration-based
car-following models to a two-dimensional force field. Depending on
the underlying car-following model, its parameterization, and optional
floor fields, it can describe manually driven or autonomous motorized
vehicles as well as bicycles driving in lane-based and non-lane-based
traffic.

Generally, the proposed model describes the directed flow
of high-speed
self-driven particles where ``high-speed'' indicates that collisions are
undesirable and kinematic aspects such as braking distance play a
significant role. This includes all of the above, and the flow of
athletes in running and 
cross-country ski Marathons, city inline-skating events, and
others~\cite{TreiberKesting-Book,TGF13-ski}. 

The paper is organized as follows. In the next two sections, the model
is specified and tested for plausibility on some standard
situations. In Section 4, the model is calibrated and validated on
trajectory data of semi-dense and congested traffic in an Indian
city. Section 5 concludes with a discussion. 

\begin{figure}
\fig{0.9\textwidth}{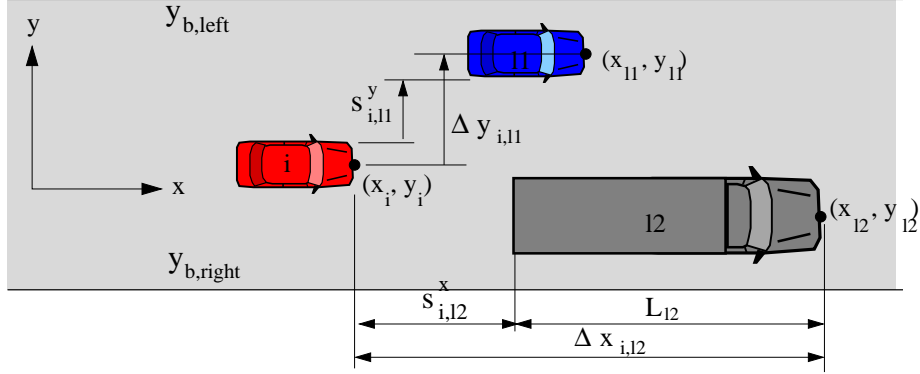}
\caption{\label{fig:sketch}Specification of the geometrical variables
  of the MTM.}
\end{figure}

\section{Specification of the Self-Driven-Particle Model}

%
While the Mixed Traffic Flow Model
(MTM) proposed in the following is designed to describe general
self-driven high-speed particles, we
will refer to them as ``vehicles'', for notational clarity.
Assuming that all considered vehicles drive in the same general
direction, we separate the motion into a longitudinal part along the
local road axis (coordinate $x$), and a transversal part perpendicular
to it (coordinate $y$). The transversal dynamics replaces the lane changing
component of conventional microscopic traffic flow models.
For reasons of simplicity, we consider
only rectangular objects (cf. Fig.~\ref{fig:sketch}).

We denote the
(front-center) position,
velocity, and social-force vectors of vehicle $i$  by
$\vec{r}=(x,y)$,
$\vec{v}=(v,w)$, and  $\vec{f}=(f,g)$, respectively.
Without loss
of generality, we assume all  masses $m_i=1$ such that the force
vectors also denote the accelerations.
The total acceleration
vector $\vec{f}=\diff{\vec{v}_i}/\!\diff{t}$ is given by a
superposition  of
several ``social forces'' according to 
\be
\label{accMTMgen}
\abl{\vec{v}_i}{t}=\vec{f}\sup{self}_i(\vec{v}_i)
 +\vec{f}\sup{int}_{i}(\vec{r}_i,\vec{v}_i, \, \{\vec{r}_j,\vec{v}_j\})
 +\sum_b \vec{f}_{ib}.
 \ee
This expression formally resembles the acceleration equation of the
social-force model for pedestrians~\cite{HelMol95}. It includes  the
self-driven acceleration $\vec{f}\sup{self}_i$, 
the interacting force $\vec{f}\sup{int}_{i}$ caused by a set $\{j\}$ of neighboring
vehicles, and the external forces $\vec{f}_{ib}$ due
to road boundaries, red traffic lights, and other constraints. Notice
that, in
contrast to~\cite{HelMol95}, we will specify these forces such that
they are consistent with the kinematic constraints and driver
anticipations that are necessary for a collision-free flow of
high-speed particles. Because we consider directed particles,
the equations of motion for the longitudinal and transversal
dynamics are qualitatively different.

\subsection{Longitudinal dynamics}

In the longitudinal direction $x$, we only consider the two-body
interaction $f_{il'}$  with the mostly interacting leader or neighbor $l'$ and
add up the boundary effects of the left and right boundaries and of
red traffic lights,
\be
\label{accMTMlong}
\abl{v_i}{t}=f_i\sup{self}+f_{il'}+\sum_b f_{ib}, \quad 
l'=\text{arg}\max_l |f_{il}|.
\ee
The reason why we only consider a single leader  is that, otherwise,
multiple leaders may lead to a 
too defensive behavior. This is particularly relevant if several
small vehicles such as motorbikes drive in parallel ahead of a big
truck. Then, it is reasonable that only the closest and/or slowest
leader should determine the truck driver's acceleration or braking reaction.

We derive the self-driven and interaction accelerations from a
conventional acceleration-based single-lane car-following (CF) model where the
acceleration
$a\sup{CF}(\Delta x_{il}, v_i, v_l)$ of the
follower $i$ is a function of 
the longitudinal distance $\Delta x_{il}=x_l-x_i$ to the leader $l$
(including the leader's vehicle length) and the
speeds $v_i$ and $v_l$ of the subject vehicle and leader,
respectively. This includes the
Intelligent-Driver Model (IDM)~\cite{Opus} and variants thereof, the
Optimal-Velocity Model (OVM)~\cite{Bando-jphys}, and the Gipps
model~\cite{Gipps81}.

In a first step, we decompose the CF acceleration $a\sup{CF}$ into a free and an
interacting part according to
\be
\label{accCF}
a\sup{CF}(\Delta x_{il}, v_i, v_l)=a\sup{CF,free}(v_i)
  +a\sup{CF,int}(\Delta x_{il},v_i,v_l), 
  \ee
where
\bea
\label{accCFfree}
a\sup{CF,free}(v_i)
  &=& a\sup{CF}(\infty, v_i, v_i), \\
\label{accCFint}
a\sup{CF,int}(\Delta x_{il}, v_i, v_l)
  &=&  a\sup{CF}(\Delta x_{il}, v_i, v_l) - a\sup{CF,free}(v_i).
\eea
Notice that, for plausible models, the interaction tends to zero when
the distance $\Delta x_{il}$ tends to
infinity  (cf.~\cite{TreiberKesting-Book} for details), 
so $a\sup{CF,free}$ can be calculated at any value for $v_l$, including
$v_i$.

In a second step, we identify the free acceleration as the self-driven component
of~\refkl{accMTMlong} and assume the interaction force $f_{il}\sup{int}$ from a
leader $l$ to be that
of the underlying car-following model whenever there is a
longitudinal overlap, i.e., an uncontrolled approach to the
leader will eventually lead to a rear-end
collision. This is the case, if the magnitude of the lateral
displacement (cf. Fig.~\ref{fig:sketch}) $\Delta y_{il}=y_l-y_i$ is less than the average
vehicle width
$\overline{W}_{il}=(W_l+W_i)/2$. Otherwise, i.e., if there is a
lateral gap $s_{il}^y=|\Delta y_{il}|-\overline{W}_{il}>0$, the interaction decreases
exponentially with this gap.
Bringing all this together, the free and vehicle interaction part  
of the longitudinal MTM acceleration~\refkl{accMTMlong} can be
expressed in terms of the CF acceleration by
\be
\label{accFree}
f_i\sup{self}=a\sup{CF,free}(v_i),
\ee
and
\be
\label{accIntLong}
f_{il'} = \alpha(\Delta y_{il'}) a\sup{CF,int}
(\Delta x_{il'},v_i,v_{l'}), \qquad l'=\text{arg}\max_l |f_{il}|,
\ee
where the attenuation factor
\be
\label{alpha}
\alpha(\Delta y_{il})=\min \left\{\exp\left(-\frac{s_{il}^y}{s_0^y}\right),
\ 1\right\}, \quad 
 s_{il}^y=|\Delta y_{il}|-\overline{W}_{il}
 \ee
contains the lateral ``attenuation scale''  $s_0^y$ as a model
parameter which can also be interpreted as a ``soft'' minimum
lateral gap (cf. Table~\ref{tab:MTM}). Notice that for sufficiently
aligned leader-follower pairs 
with a lateral overlap, $s_{il}^y<0$, the expressions~\refkl{accFree}
and~\refkl{accIntLong} reduce to the
CF acceleration of the immediate (mostly interacting)
leader, i.e., the model reverts to the underlying CF model.
To ensure that
situations with a longitudinal overlap, i.e., a negative longitudinal gap $s_{il}=x_l-x_i-L_l<0$ (where
$L_l$ is the length  of
the leader), leads to a consistent behavior, the CF model needs to
return its maximum 
deceleration $b\sub{max}$ (emergency braking)  whenever its distance
argument $\Delta x_{il}$ is less than $L_l$ (i.e., the gap
$s_{il}$ is negative).

The external forces $f_{ib}$ can come from road
boundaries, red traffic lights and other constraints such as stop
signs. A red traffic light or stop sign is modelled as a very short and very
wide standing virtual
vehicle occupying the complete extension of the stopping line. Notice
that these virtual vehicles are considered separately from the real
leader, i.e., a follower may brake even if the leader decides to pass
the traffic light about to go red. Road
boundaries are modelled by decelerating ``viscous'' shear forces
whose influence increases with decreasing lateral gap
$s_{ib}^y=\pm(y_b-y_i)-W_i/2$ between vehicle and 
road boundary (plus/minus for the right/left boundary, respectively)
and becomes equal to $-b_b$ if the vehicle touches the road
boundary,
\be
\label{accBoundary}
f_{ib}=- b_b \left(\frac{v_i}{v_0}\right) 
\exp \left(-\frac{s_{ib}^y}{s_{0b}^y}\right).
\ee
Notice that~\refkl{accBoundary} is essentially the shear force
resulting from~\refkl{accIntLong} with~\refkl{alpha} when representing
the road boundary by a series of standing virtual vehicles just
outside of the road. However, since leaving the designated road
surface is typically less harmful than colliding with standing vehicles
(at least, if the road has a paved
emergency strip), the lateral boundary scale $s_{0b}^y$ should be
smaller than $s_0^y$ and the deceleration $-b_b$ when the wheels just
touch the boundary is significantly lower than $b\sub{max}$. 
Furthermore, ~\refkl{accBoundary} contains another
softening factor $\frac{v_i}{v_0}$ allowing to ``squeeze'' through very
narrow bottlenecks at low speeds or partially leave the road when
nearly standing. Generally, both 
road sides exert shear forces  according
to~\refkl{accBoundary}. For very narrow roads, this reduces the speed
of the vehicle considerably, even without obstructions from other vehicles.

\subsection{Lateral dynamics}
%
According to the general equation~\refkl{accMTMgen}, the lateral
social force or acceleration
 $g_i=\abl{w_i}{t}$ of vehicle $i$ is 
composed of an external tactical component, road boundary constraints,
and a traffic-dependent component,
\be
\label{accMTMlat}
g_i=g_i\sup{self}+g_i\sup{b}+g_i\sup{int}.
\ee

The \emph{tactical} component $g_i\sup{self}$ represents a mandatory or
anticipatory lateral displacement (corresponding to lane changes in
lane-based traffic) when 
entering or leaving a road, passing a bottleneck,  or simply obeying
``drive to the right'' or ``drive to the left'' regulations. Its
magnitude (of  the order of $\unit[1]{m/s^2}$) depends on the 
urgency of the action. It will  
not be considered in this paper.

The  \emph{road boundary component} has the same functional lateral
dependency as the longitudinal component,
\be
\label{gb}
g_i\sup{b}
=\pm \tilde{b}_b \left(\frac{v_i}{v_0}\right) 
\exp \left(-\frac{s_{ib}^y}{\tilde{s}_{0b}^y}\right),
\ee
where both boundaries are superposed and the ``+'' sign applies for
the left boundary ($y$ is increasing 
to the right). 
However, since drivers avoid the road boundary mainly by steering and
not by braking, both the magnitude $\tilde{b}_b$ 
of the lateral
acceleration when touching the boundary, and the lateral decay
 scale $\tilde{s}_{0b}$ are larger than that for the longitudinal
 force.

The \emph{traffic-dependent interaction component} $g_i\sup{int}$ is relevant if
a driver is obstructed by slower leading vehicles or by vehicles
driving closely at the side. Depending on the parameterization, it
also reflects to ``look back'' before 
 initiating right or left movements and perform these maneuvers
 only if there is no
accident risk and no back vehicles are obstructed unreasonably.  
In our approach, the lateral traffic
interaction is mediated by a simple generalized ``car-following
model'' for the 
lateral coordinates whose ``desired lateral speed'' reflects the
social forces and depends on the
surrounding vehicles. However, using a fully fledged CF model for the lateral
dynamics (e.g., that used for the longitudinal dynamics) is
problematic since CF models only give useful results for 
forward moving vehicles while lateral motion is
symmetric. Particularly, the actual and desired lateral speeds may be negative
reflecting a (desire to) move to the left. Moreover,
this approach would entail several unnecessary new model parameters.   

We therefore consider the simplest model (i) coping with negative speeds
and displacements, (ii) taking into account the relevant factors,
namely the subject's lateral speed $w_i$, the lateral
speed $w_j$ of the interacting vehicles $j$ and their lateral
displacement $\Delta y=y_j-y_i$, and (iii) fulfilling the plausibility
criteria listed in Chapter 11.1 of~\cite{TreiberKesting-Book}.
Specifically, we consider an OV-like  model 
whose desired lateral speed $w^0_i$  is mediated by the social
forces discussed below,
\be
\label{accIntLat}
g_i\sup{int}=\abl{w_i\sup{int}}{t}=\frac{w^0_i-w_i}{\tau}.
\ee
The model parameter $\tau$  denotes the speed adaptation time
(cf. Table~\ref{tab:MTM}). Furthermore, the heading of the vehicles
is restricted to a cone of angle $2\theta$ parallel to the
longitudinal direction, i.e., the ratio $w/v$ is restricted by $\pm
\tan \theta$. By formulating the lateral acceleration in terms
of the OV-like model~\refkl{accIntLat} with the desired lateral speed
depending on the interactions rather than formulating the interaction
directly, we take into account that the lateral social
force does not only depend on the lateral gap to the neighbors but
also on the lateral speeds of the considered and interacting vehicles. 

Since $w^0_i=0$ (corresponding to ``drive ahead'') without
vehicle-vehicle interaction, we can interpret 
the contribution $-w_i/\tau$ as the ``free'' lateral acceleration
while the lateral desired speed $w^0_i$ includes the contributions of
all relevant neighbors,
\be
\label{w0i}
w^0_i=\sum_j w^0_{ij}.
\ee
Notice that, in contrast to the longitudinal interaction, all
neighbors with a relevant interaction are included in order 
to avoid accidents. 

We specify the contribution
$w^0_{ij}$ of an isolated neighboring vehicle $j$ by generalizing the
lane-changing model   
MOBIL~\cite{MOBIL-TRR07} to a continuous lateral coordinate. 
MOBIL states that the
\emph{incentive} to change lanes is proportional to the difference between
the potential longitudinal acceleration at the new lane and the actual acceleration
at the old lane. In the present continuous model, the
difference is replaced by the partial derivative of the longitudinal
acceleration $f_{ij}$ with
respect to $y$ (acceleration shear), so $w^0_{ij} \propto \partial
f_{ij}(x,y)/\partial y$ with $f_{ij}(x,y)$ according
to~\refkl{accIntLong}.  However, for  
small lateral displacements with an overlap 
($|\Delta y|<\overline{W}_{ij}$ or $s_{ij}^y<0$), this gradient
is zero according  and a modification is
necessary to avoid getting stuck behind a slower vehicle. 
Assuming that, for this case, 
the lateral desired speed 
is proportional to the displacement and that it is a continuous
function everywhere,  we arrive at following expression for the desired
velocity component induced by neighbor $j$:
\be
\label{vDesTrans}
w^0_{ij}= \twoCases{\lambda s_0^y\ablpart{f_{ij}}{y}} 
   {\text{ if } |\Delta y_{ij}| \ge \overline{W}_{ij}}
   {\lambda f_{ij} \frac{\Delta y_{ij}}{\overline{W}_{ij}}}
   {\text{ if } |\Delta y_{ij}| < \overline{W}_{ij}}.
\ee
The new model parameter $\lambda$ (cf. Table~\ref{tab:MTM}) denotes
the sensitivity for 
lateral motion. Specifically, $\lambda s_0^y$ indicates the lateral
desired speed induced by a longitudinal acceleration shear of
$\unit[1]{m/s^2}$ per meter, and $\lambda$ itself the maximum ratio
between the desired lateral speed and the longitudinal interaction
deceleration induced by vehicle $j$.  However, expression~\refkl{vDesTrans}
still lacks a dependence on the relative lateral speed $w_j-w_i$ which
clearly is relevant in order to avoid collisions or prevent unnecessary
steering actions. Since such a term cannot be derived from MOBIL, we
augment~\refkl{vDesTrans} in the simplest possible way that satisfies
the plausibility condition ``no lateral force if there is no
longitudinal force'' by adding to~\refkl{vDesTrans} a linear relative
speed dependence  
$-\lambda_{\Delta w}\text{sign}(\Delta y_{ij})(w_j-w_i)$ multiplied
  by~\refkl{vDesTrans} itself. The parameter
$\lambda_{\Delta w}$ denotes the sensitivity to lateral speed
differences (cf. Table~\ref{tab:MTM}) and should satisfy
$\lambda_{\Delta w}|w_j-w_i|<1$ for all reasonable relative lateral
speeds in order to prevent an implausible driving behavior
(lateral ``drag-along'' effect).
Performing the partial derivative on~\refkl{accIntLong} and including
the above factor, we obtain following
explicit expression for the pair interaction,
\be
\label{w0ij}
w^0_{ij}=\lambda \tilde{\alpha}(\Delta y_{ij}) a\sup{CF,int}_{ij} \,
\left[1-\lambda_{\Delta w} (w_j-w_i) \text{sign}(\Delta y_{ij}) \right],
\ee
where $a\sup{CF,int}_{ij}<0$ is the interacting part of the
(longitudinal) CF acceleration from
leader $j$, and the lateral attenuation factor is given by
\be
\label{tilalpha}
\tilde{\alpha}(\Delta y_{ij})=
 \text{sign}(\Delta y_{ij}) \twoCases
{1+s_{ij}^y/\overline{W}_{ij}}{s_{ij}^y<0}
{\exp\left(-s_{ij}^y/\tilde{s}_0^y\right)}{s_{ij}^y\ge 0}, 
\quad s_{il}^y=|\Delta y_{il}|-\overline{W}_{il}.
\ee
Since drivers prefer steering to braking in order to avoid a collision
with another vehicle, we have introduced a different lateral interaction scale
$\tilde{s}_0^y$ which is larger than its counterpart $s_0^y$ for the
longitudinal accelerations.

The lateral forces are repulsive and increase linearly with the
lateral distance in case of a lateral overlap while they decay
exponentially, otherwise. Notice that, for the non-overlapping case
$s_{ij}^y>0$ and zero lateral speeds $w_i$ and $w_j$, the ratio of the
longitudinal and lateral repulsive forces (accelerations) 
from a leading vehicle is given by the
dimensionless constant $\lambda/\tau$. 

Expression~\refkl{w0ij} is only valid for leading neighbors. In order
to include followers, 
\bi
\item the role of the leader and follower has to be swapped, i.e., the
  lateral interaction is calculated for the follower,
\item the sign of the contribution has to be swapped, reflecting
  \emph{actio=reactio}. 
\ei
In terms of a generalized lane-changing model
MOBIL~\cite{MOBIL-TRR07}, including only the leaders in~\refkl{w0i}  
corresponds to a \emph{politeness factor} $p=0$ and no active safety
criterion, while including all
neighbors corresponds to $p=1$. Obviously, general values of $p$ are
realized by weighting the followers with $p$. For an effective
implementation, only leaders and followers are included whose maximum CF
interactions exceed  a certain threshold,
$|a\sup{CF,int}_{ij}|>a\sub{thr}$ and
$|a\sup{CF,int}_{ji}|>a\sub{thr}$, respectively.

Finally, a safety criterion similar to that of MOBIL
 is automatically included provided
that $p>0$. This is ensured since we require from the underlying car-following
models to return $a\sup{CF,int}=-b\sub{max}$
whenever the longitudinal gap $s<0$. This means,
the above expressions give a 
maximum repulsive lateral interaction if neighbors overlap
longitudinally with the subject vehicle, i.e. drive nearly in
parallel or following very closely or at high speed.

In summary, the complete MTM model is defined by the longitudinal
dynamics~\refkl{accMTMlong} with~\refkl{accIntLong}, \refkl{alpha},
\refkl{accBoundary} and the lateral dynamics~\refkl{accMTMlat} with
\refkl{gb}, \refkl{accIntLat}, \refkl{w0i}, \refkl{w0ij}, and
\refkl{tilalpha}. Typical values of its nine parameters are listed in
Table~\ref{tab:MTM}. In addition, one needs an underlying
CF model whose acceleration output  is separable into a free-flow and
interacting part.


\section{Model Visualization and Plausibility Tests}

To visualize the model and test whether it produces plausible driver
actions and trajectories, we need to fully specify the model by
providing an underlying car-following model. We
investigated several well-known models including the
Intelligent-Driver Model (IDM)~\cite{Opus}, one of its derivatives, the
``Adaptive Cruise Control'' (ACC) model)~\cite{kesting-acc-roysoc,TreiberKesting-Book},
the optimal-velocity model
(OVM) of Bando et al.~\cite{Bando-jphys},  and the Gipps
model~\cite{Gipps81} in the simplified version presented
in~\cite{TreiberKesting-Book}. The ACC model can serve as a prototype
for the longitudinal control of autonomous vehicles. Unlike the
original IDM, it has a triangular fundamental diagram and introduces an
additional "coolness factor" preventing the IDM's unrealistically
strong driver response to cut-in maneuvers that regularly happen in
mixed traffic, so we will use it as the main underlying model in the
calibration and validation in Sec.~\ref{sec:calVal}. 
The remaining dynamical aspects and the parameter set
is the same as in the IDM.  The OVM is qualitatively different from the
ACC and Gipps models since it does not depend on the leader's
speed. Its natural extension, the Full Velocity Difference
Model (FVDM) of Jiang et al~\cite{Jiang-vDiff01}, does not satisfy the basic
plausibility conditions of car-following in free and congested traffic
(cf. Chapter 11.1 of~\cite{TreiberKesting-Book}). This prevents a
decomposition into free and interacting parts according
to~\refkl{accCFfree} and~\refkl{accCFint}, so this model is not
applicable. 

For reference, we use the OVM in the form given in
(cf. Chapter 10 of~\cite{TreiberKesting-Book})

\begin{equation}
\label{OVM}
  f^{\rm OVM}(s,v,v_l)=\frac{v_{\rm opt}(s)-v}{\tau}, \quad v_{\rm
    opt}(s)=v_0\frac{\tanh\left(\frac{s}{\Delta s}-\beta\right)+\tanh
    \beta}{1+\tanh \beta}.
\end{equation}

\begin{figure}
\fig{1.0\textwidth}{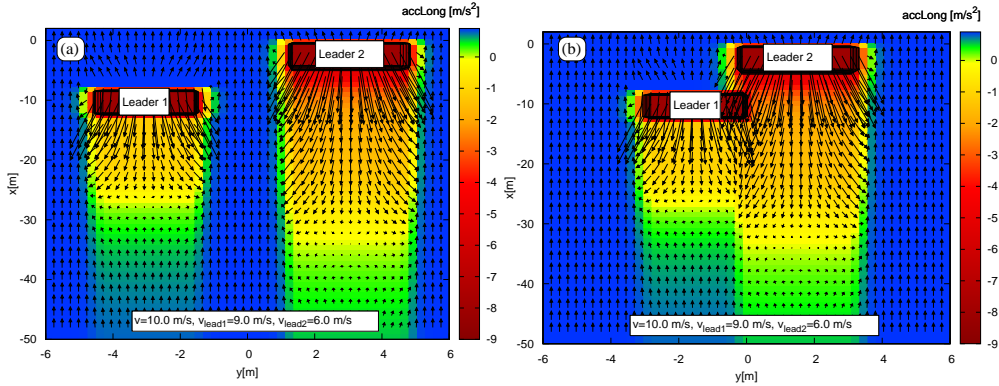}
\caption{\label{fig:accField-IDM}Acceleration vector field of the proposed
  model for a following vehicle (actual speed $v=\unit[10]{m/s}$, desired
  speed $v_0=\unit[18]{m/s}$) at arbitrary positions $\vec{r} = (x,
  y)$ attempting to 
  pass two slower leaders whose outlines are given by the white
  boxes. Additionally, the longitudinal component is given by the
  color-coded background. (a) passing between the
  leaders is possible; (b) the follower needs to circumvent the two
  leaders. The underlying car-following (CF) model is the ACC model
  (modified IDM,~\cite{kesting-acc-roysoc}) 
  with parameters $T$, $s_0$, $a$ and $b$ given in
Table~\ref{tab:CFcalibr} for cars. The  MTM parameters
of the generic MTM particle model are given in Table~\ref{tab:MTM}.
}
\end{figure}

\begin{figure}
\fig{1.0\textwidth}{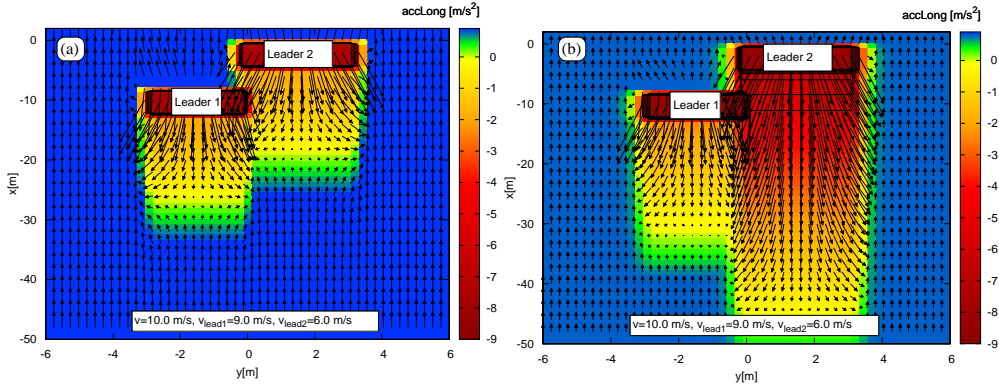}
\caption{\label{fig:accField-OVM-Gipps}Acceleration vector field as in
  Fig.~\ref{fig:accField-IDM} (b) for other underlying CF
  models. (a) OVM~\cite{Bando-jphys}, Eq.~\refkl{OVM}; (b) simplified Gipps model
  as defined in~\cite{TreiberKesting-Book}. In both models, the
  desired speed $v_0=\unit[18]{m/s}$. The OVM interaction
  parameters are $\tau=\unit[5]{s}$, $\beta=1$, and $\Delta
  s=\unit[12]{m}$. The remaining Gipps model parameters $\Delta t=T$,
  $a$, and $b$ are given in
Table~\ref{tab:CFcalibr} for cars. The  MTM parameters are given in
Table~\ref{tab:MTM}. 
}
\end{figure}

The Figures~\ref{fig:accField-IDM} and~\ref{fig:accField-OVM-Gipps}
show the acceleration vector field of 
the MTM  as a function 
of the position $(x,y)$ of the subject vehicle relative to that of two adjacent
vehicles. The underlying CF models are the ACC model, the OVM, and the
simplified Gipps model in the Figures 
\ref{fig:accField-IDM}, \ref{fig:accField-OVM-Gipps}(a)), and
\ref{fig:accField-OVM-Gipps}(b), respectively. Since the speeds of the
other vehicles are significantly below the subject driver's
desired speed $v_0=\unit[20]{m/s}$, the driver has an incentive to
get past them. The MTM acceleration fields
reflect following common plausible actions: 

\begin{itemize}

\item If the other vehicles are leaders and sufficiently far away, or
  if they are on either side with sufficient lateral space, the acceleration is
  essentially that on a free road, i.e., positive and in the
  longitudinal direction.

\item in case of a single-lane string of vehicles, the longitudinal
  part of the MTM reverts to
  the underlying CF model inheriting all of its properties.
  particularly, the MTM based on the IDM or the ACC and Gipps models
  is crash free while that based on the OVM is not.

\item If the subject vehicle approaches slower leaders and an action is
necessary, the situation is handled by simultaneously decelerating and
steering. Depending on the relative position, the acceleration or the
steering component prevails. 

\item If a passage between the leaders is possible, the subject
  vehicle steers towards the center of the lateral gap and
  overtakes. Otherwise, the 
  group of  the two leaders is circumvented to the right or
  left. There is also the theoretical possibility that, in a
  homogeneous stationary situation, a follower gets
  stuck between leaders driving close together (in 
  Fig.~\ref{fig:accField-IDM}(b), it is at $x=-\unit[28]{m}$ and
  $y=-\unit[0.7]{m}$, and in Fig.~\ref{fig:accField-OVM-Gipps} it is
  at similar locations). However, the dynamical
  simulation shows that, in reality, such situations are of a short
  duration, if they occur at all.
\end{itemize}
Comparing the underlying models reveals a difference, however:
although Leader~2 is further ahead, its braking and steering effect on
the follower is stronger compared to that
of Leader~1 for the IDM/ACC/Gipps models. The reason is that Leader~2 is
slower than Leader~1 and these CF
models depend on relative 
longitudinal speeds. The MTM inherits this property.   In contrast,
the situation is reversed for the OVM where only gaps
matter. Generalizing this, all behavioral aspects (and anticipations)
of the underlying CF model carry over to the acceleration, braking and steering
strategy of the MTM.

\begin{figure}
\fig{0.9\textwidth}{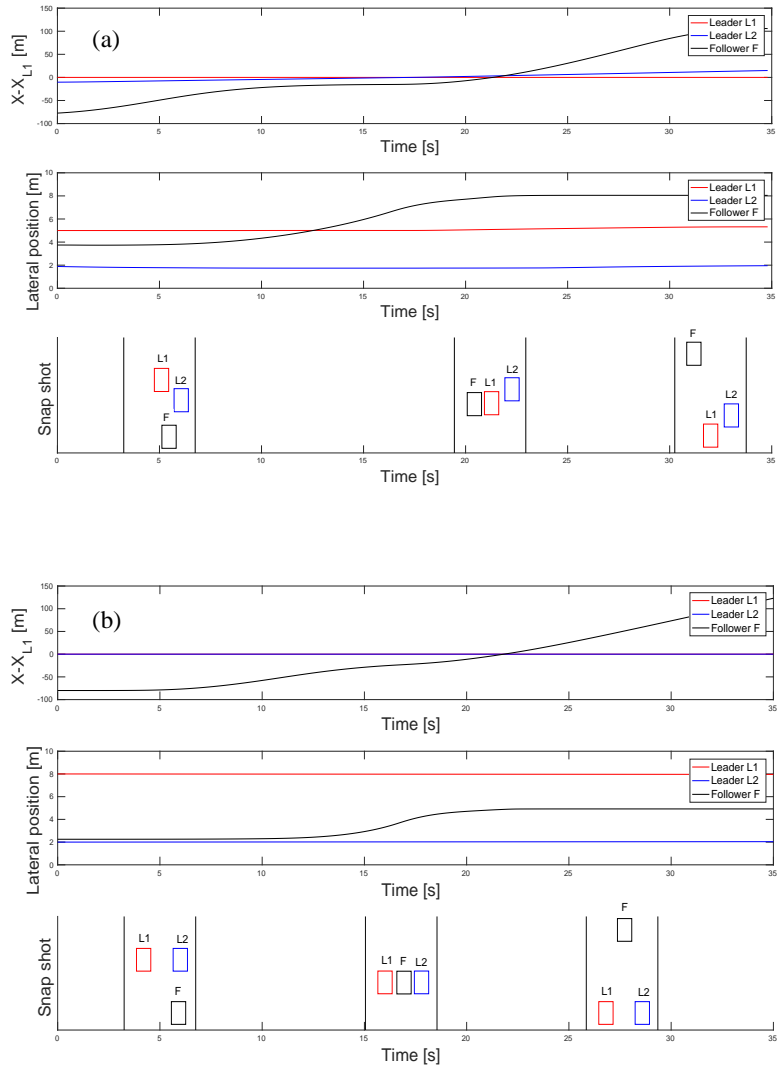}
\caption{\label{fig:timeseries}Time series for two passing
  maneuvers. (a), A follower F passes between two slower leaders 
L1 and L2. (b), The follower circumvents a group of two slower leaders
to the left. Each of the two 
panels display the longitudinal distance to Leader L1 (top), the
lateral position relative to the road axis 
(middle), and snapshots of the configuration (bottom).}
\end{figure}

As a further plausibility test, we have simulated, for the IDM/ACC as
underlying model, three idealized cases: 1. Passing of an isolated
slower vehicle, 2. passing between two slower leaders
(Fig.~\ref{fig:timeseries}(a)),  and
3. Circumventing a group of two slower leaders in order to pass
(Fig.~\ref{fig:timeseries}(b)). The simulated drivers display a plausible behavior:
first, they move with a lateral component in order to find a suitable
gap, then they accelerate to pass the slower vehicles. Notice that the
vehicles L1 and L2 hardly react to Vehicle F once it becomes their
leader. This plausible behavior is, again, a consequence of the
relative speed dependence of the IDM/ACC model producing only little
interaction if the leader is faster than the follower. Using the OVM
would lead to implausible (too strong) reactions to L1 and L2.

\section{\label{sec:calVal}Calibration and Validation} 

\begin{figure}
\fig{0.9\textwidth}{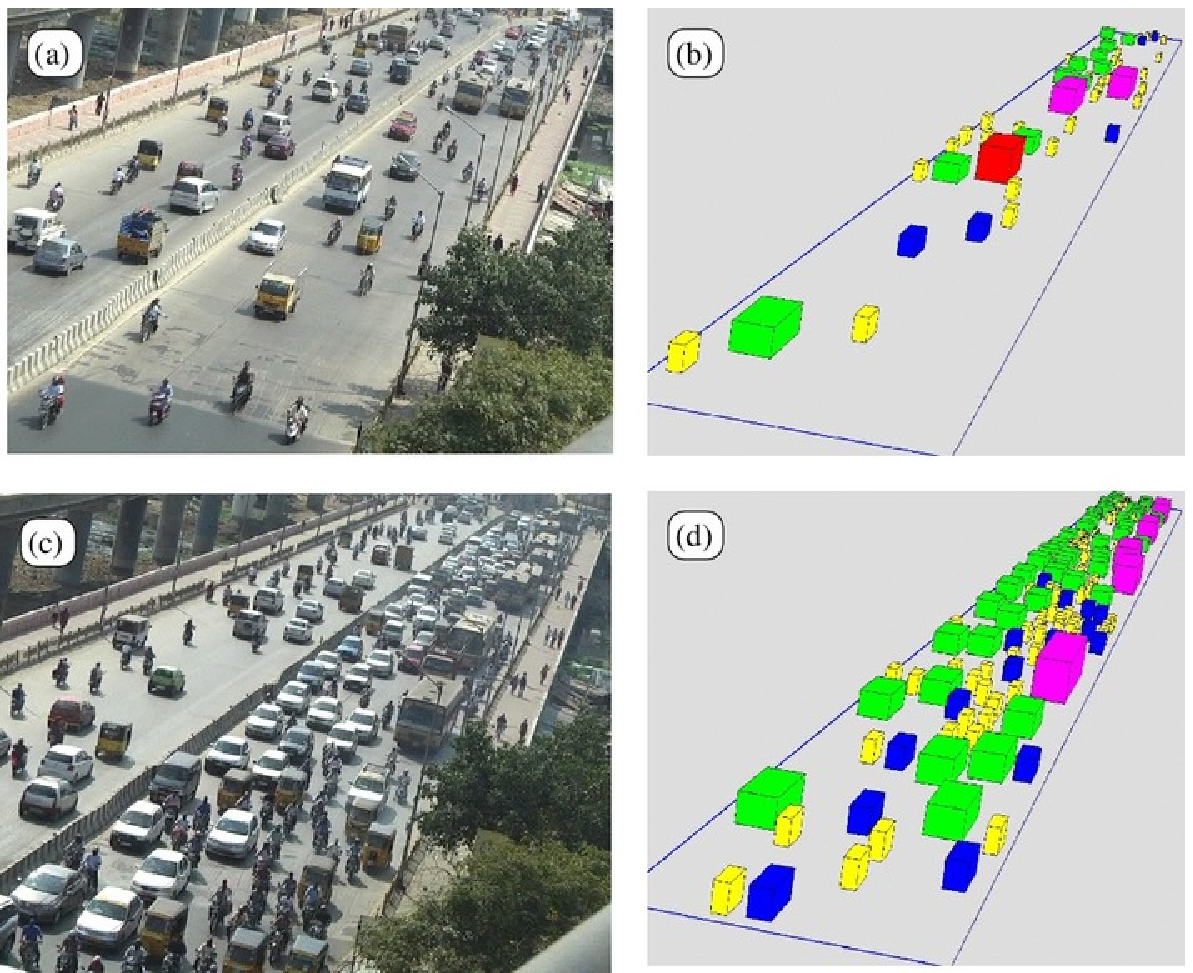}
\caption{\label{fig:photoSimTraffic}Snapshots and Simulation of mixed
  traffic. (a), Medium to dense traffic on a \unit[250]{m} road stretch 
in Chennai, India. (b), Simulation of this situation with initial
conditions taken from the observations. 
(c)-(d), Observed and simulated congested mixed traffic on the same
road. Videos for all four panels are available.}
\end{figure}

\begin{table}
  \caption{\label{tab:CFcalibr}Calibrated ACC model (IDM) parameters for
    the vehicle mix of the observations shown in
    Fig.~\ref{fig:photoSimTraffic}. The maximum braking deceleration
    has always been set to $b\sub{max}=\unit[9]{m/s^2}$
  }

  \begin{center}
  \begin{tabular}{lccccccc}\hline
  Vehicle & Length & Width  & $v_0$  & $T$ & $s_0$
   & $a$ & $b$  \\
  type & [m] & [m] & range [m/s] & [s] & [m]  & \unit{[m/s$^2$]}
   & \unit{[m/s$^2$]} \\ \hline
  Motorcycle    & 1.8  & 0.6 & 25-18 & 0.3 & 0.5 & 2.0 & 2.0\\
  Car           & 4.2  & 1.7 & 18-12 & 0.8 & 2.0 & 1.0 & 1.0\\
  Bus           & 10.3 & 2.1 & 14-10 & 1.0 & 2.0 & 1.0 & 1.0\\
  Auto-Rickshaw & 2.6  & 0.9 & 5-6   & 1.0 & 2.0 & 1.0 & 1.0\\ \hline
  \end{tabular}
  \end{center}
\end{table}

\begin{table}
  \caption{\label{tab:MTM}Calibrated Parameters of the mixed traffic flow model}

  \begin{center}
  \begin{tabular}{lr}\hline
    Parameter & value \\ \hline
    maximum angle to road axis $\theta$ & \unit[0.2]{rad} \\
    lateral interaction scale for braking $s_0^y$ &\unit[0.15]{m}\\
    lateral interaction scale for steering $\tilde{s}_0^y$ &\unit[0.30]{m}\\
    boundary interaction scale for braking $s_{0b}^y$ &\unit[0.15]{m} \\
    boundary interaction scale for steering $\tilde{s}_{0b}^y$ 
             &\unit[0.25]{m} \\
    lateral sensitivity $\lambda$   & \unit[0.4]{s} \\
    lateral time constant $\tau$  & \unit[1]{s} \\ 
    sensitivity to lateral relative speeds $\lambda_{\Delta w}$ 
             & \unit[0.7]{s/m} \\
    politeness factor $p$  & 0.2
\\ \hline
  \end{tabular}
  \end{center}
\end{table}

\begin{table}
  \caption{\label{tab:calVal}Observed and simulated
    characteristics of the mixed traffic flow at the test site in
    Chennai, India (Fig.~\ref{fig:photoSimTraffic})}

  \begin{center} 
\begin{tabular}{lrr}\hline
Variables                       & Observed & Simulated \\ \hline
\multicolumn{3}{c}{Calibration-Medium}                 \\ \hline
Average travel time [s]  & 7.24     & 6.16      \\
Travel time standard deviation [s] & 3.02     & 3.90      \\
Entry flow[veh/s]        & 2.79     & 2.42      \\
Exit flow[veh/s]         & 1.96     & 2.21      \\
Number of lateral shifts & 11.00    & 14.00     \\ \hline
\multicolumn{3}{c}{Validation-Congested}               \\ \hline
Average travel time [s]  & 12.82    & 13.41     \\
Travel time standard deviation [s] & 7.28     & 5.95      \\
Entry flow[veh/s]        & 1.87     & 1.79      \\
Exit flow[veh/s]         & 2.31     & 2.68      \\
Number of lateral shifts & 14.00    & 9.00      \\ \hline
\end{tabular}
  \end{center}
\end{table}

In this section, we test the predictive power of the proposed MTM on
observed mixed traffic flow in the city Chennai, India. The Panels (a)
and (c) of Fig.~\ref{fig:photoSimTraffic}
depict the situation: medium and congested traffic
on a 250 m long homogeneous section of a divided urban road with
nominal six lanes per roadway. Each roadway has a width of
\unit[12]{m}.
See also Supplemental
Material Videos 1-4. We only consider the driving direction towards the observer
(notice that there is left-hand traffic in India). 

In order to calibrate and validate the MTM, we have simulated a \unit[1\,500]{m} long
homogeneous section with properties as that of the observation site
using the ACC model as underlying 
car-following model. Initially, the first \unit[250]{m} (observation
section) are populated
with the observed vehicle positions and types (motorcycles, cars,
busses, and auto-rickshaws) depicted in
Fig.~\ref{fig:photoSimTraffic}(a) or
Fig.~\ref{fig:photoSimTraffic}(c), respectively,  and with initial
velocities as derived from
the videos. Notice that there were no
trucks at the time of recording. From the videos, we have also
inferred the typical vehicle dimensions and approximate desired speed
distributions (Columns 2-4 of Table~\ref{tab:CFcalibr}).
For illustrative purposes, the 1d partial densities
obtained from the video were 170 motorcycles/km, 55
cars/km, 10 buses/km, 15 auto-rickshaws/km for medium traffic, 
and 450 motorcycles/km, 225 cars/km, 35 buses/km, and 80
auto-rickshaws/km for congested traffic. The rest of the simulated
road section is initialized by 
repeating this configuration until the end of the road is
reached.

During the simulation (simulation time \unit[20]{s}), no additional
vehicles were introduced at 
the upstream boundary while the downstream boundary is free, i.e.,
vehicles are removed once they cross the downstream boundary resulting in
free traffic for the immediate followers. The test section for calibration and validation is in the range
$x \in [\unit[180]{m}, \unit[227]{m}]$. The road section was long
enough such that, during the simulation time,  only
vehicles  defined by the initial conditions could enter the test
region, and also traffic waves near the downstream boundary cannot
reach the test region, i.e., the boundary conditions do
not play a role. 

We have calibrated the model on medium traffic
(Fig.~\ref{fig:photoSimTraffic}~(a)) by minimizing a mixed objective
function consisting of the 
observed macroscopic characteristics of traffic flow listed in Table~\ref{tab:calVal},
\be
\label{objFun}
S(\vecbeta)=\sqrt{\sum_{i=1}^5\left(
  \frac{X_i\sup{sim}(\vecbeta)-X_i\sup{obs}}{X_i\sup{obs}}\right)^2},
\ee
where $X_i$ denotes average travel time through the test section, its
standard deviation, the average entry and exit flows into and from the
test section, and the total number of ``lateral shifts'' in this region.  
A lateral shift event happens if (i) the vehicle moves laterally by
more than \unit[1]{m} in the same direction, and (ii) its leader as
defined by~\refkl{accIntLong} changes. The optimization is performed
by MATLAB's standard genetic algorithm.
Figure~\ref{fig:photoSimTraffic}~(b) shows a snapshot of the simulation for medium traffic after
the calibration The estimated model parameters are given in the Tables~\ref{tab:CFcalibr} and~\ref{tab:MTM}.
We observed that all simulated flow characteristics of the
calibrated model agree with that of the data to within \unit[20]{\%} 

In order to validate the model, we have applied the calibrated model
to the congested situation (Figs~\ref{fig:photoSimTraffic}~(c) and
(d)). With respect to the 
calibration, we have only changed the initial vehicle configuration
to the observed one keeping all parameters to
their calibrated values. Apart from the number of lateral shifts, the
other simulated criteria differ less than \unit[20]{\%}  from the
observations (Table~\ref{tab:calVal}). In view of the fact that the validation
scenario is qualitatively different (congested rather than medium
traffic), this indicates a good predictive power.  The difference in
the number of lateral shifts is possibly explained by a more aggressive
driving behavior in congested traffic, or by additional cooperation,
which is not reflected by the model.

\section{Concluding Remarks}

We have proposed a two-dimensional time-continuous model for mixed
traffic flow of motorized and non-motorized vehicles, the mixed traffic
flow model (MTM), and tested it on
real observations. The MTM generalizes a conventional
acceleration-based single-lane car-following (CF) model to two
continuous dimensions. Because it is based on a
CF model, the MTM 
 can be considered as a framework rather than a specific model. 
This means that all behavioral aspects of the underlying CF
model carry over to the two-dimensional model: acceleration and
steering are intrinsically connected. Methodically, the proposed
approach is similar to that leading to the discrete lane-changing
model MOBIL~\cite{MOBIL-TRR07}. In fact, when introducing additional
thresholds or ``floor fields'' 
representing lane markers, the model reverts to an
integrated car-following and lane-changing model which, unlike MOBIL,
models explicit steering maneuvers and also contains, for a politeness
factor $p>0$, cooperative
elements such as zipper merging. This will be an interesting research
topic for the future. While the MTM is formally similar to
the social-force model for pedestrians~\cite{HelMol95}, its basic
assumptions (explicit consideration of vehicle kinematics,
unidirectional flow, no collisions if the CF model is collision free)
are markedly different.  We expect that the MTM
can serve as a general model platform for ``high-speed'' self-driven
particles and can be parameterized or extended to describe bicycle traffic,
acceleration and steering of autonomous vehicles (Markoff, 2010), and
crowd flow in mass-sports events (Treiber et al., 2015). 

\section*{Acknowledgements}  The first author is supported by a fellowship of the Alexander von Humboldt Research Foundation, Germany at the Technical University of Dresden, Germany.


\bibliographystyle{plainnat}
\bibliography{database}

\end{document}